\documentclass[conference]{IEEEtran}
\IEEEoverridecommandlockouts
\usepackage{amsmath,amssymb,amsfonts}
\usepackage{algorithmic}
\usepackage{graphicx}
\usepackage{dblfloatfix}
\usepackage{textcomp}
\usepackage[nolist]{acronym}
\usepackage{textcomp}
\usepackage{booktabs}
\usepackage{tabularx}
\usepackage[utf8]{inputenc}   
\usepackage{textgreek}
\usepackage{algorithm}
\usepackage{orcidlink}
\usepackage{tikz}
\usepackage{pifont}
\usepackage{tabularx}
\usepackage{framed} 
\usepackage{listings} 
\usepackage{array} 
\usetikzlibrary{positioning, arrows.meta}
\usepackage[table]{xcolor}
\usepackage[font=small,skip=0pt]{caption}

\newcommand{\CL}[1]{\textbf{\textcolor{red}{CL: #1}}}

\usepackage{fancyhdr}
\renewcommand{\CL}[1]{#1}

\usepackage{booktabs}
\usepackage[backend=biber,style=numeric]{biblatex}

\definecolor{LightGray}{gray}{0.9}
    
\def\BibTeX{{\rm B\kern-.05em{\sc i\kern-.025em b}\kern-.08em
    T\kern-.1667em\lower.7ex\hbox{E}\kern-.125emX}}

\begin{document} 

\title{The Economics of Autonomy: Real-Time Risk Indexing for Insurable AI-Driven 6G Systems
}

\author{   
    \IEEEauthorblockN{Anthony~Kiggundu\orcidlink{0000-0003-3921-4260}\IEEEauthorrefmark{1}\IEEEauthorrefmark{2},~Michael~Zentarra\orcidlink{0009-0001-7829-0058}\IEEEauthorrefmark{1},~Christoph Lipps\orcidlink{0000-0001-6382-9156}\IEEEauthorrefmark{1}~and~Hans~D.~Schotten\orcidlink{0000-0001-5005-3635}\IEEEauthorrefmark{1}\IEEEauthorrefmark{2}}
	\IEEEauthorblockA{
		\IEEEauthorrefmark{1}German Research Center for Artificial Intelligence (DFKI), Germany\\
		\IEEEauthorrefmark{2}RPTU University of Kaiserslautern-Landau, Germany\\
	}
}

\maketitle
\thispagestyle{fancy}

\begin{abstract}
The transition to sixth-generation (6G) networks transforms wireless infrastructure into a cognitive substrate supporting \CL{Vehicle-to-Everything (V2X), Industrial IoT (IIoT), and Integrated Sensing and Communication (ISAC)}. In this paradigm, autonomous \CL{agentic AI} performs orchestration at millisecond scales, rendering traditional static governance frameworks fundamentally inadequate for risk management. This \CL{paper} introduces \CL{GIRAF(Governance-Integrated Risk and Assurance Framework)}, a Governance-as-Code (GaC) framework for real-time risk quantification and trust modulation in agentic 6G systems. GIRAF derives a continuous Aggregate Risk Index ($R_{t}$) from machine-readable runtime signals, including epistemic confidence, network jitter, and verification latency.

A core contribution is the formalization of the verification–staleness trade-off, where safety mechanisms induce risk if computational latency exceeds 6G deadlines. \CL{We} demonstrate that GIRAF identifies 'Confidence Gaps'—discrepancies between agent-reported certainty and environmental ground-truth, triggering automated safety envelopes when conditions deteriorate. Crucially, GIRAF serves as the foundational governance groundwork and conceptual 'glue' that externalizes these technical risks into machine-readable telemetry. Through simulations with fine-tuned Large Language Models (LLMs), we validate that the framework preserves operational integrity while providing the essential actuarial baseline required for multi-stakeholder liability attribution and dynamic premium quantification in the 6G ecosystem.

\end{abstract}


\newacro{5g}[5G]{Fifth Generation}
\newacro{6g}[6G]{Sixth Generation}
\newacro{as}[AS]{Authentication Server}
\newacro{atm}[ATM]{Automated Teller Machine}
\newacro{ai}[AI]{Artificial Intelligence}
\newacroplural{ais}[AIs]{Artificial Intelligences}
\newacro{b5g}[B5G]{Beyond 5G}
\newacro{ban}[BAN]{Body Area Network}
\newacro{bsi}[\textit{BSI}]{\textit{Federal Office for Information Security}}
\newacro{bdr}[BDR]{Bit Disagreement Rate}
\newacro{bs}[BS]{Base Station}
\newacro{ca}[CA]{Certification Authority}
\newacro{cav}[CAV]{Connected Autonomous Vehicles}
\newacro{cc}[CC]{Common Criteria}
\newacro{cir}[CIR]{Channel Impulse Response}
\newacro{cr}[CR]{Challenge-Response}
\newacro{cpu}[CPU]{Central Processing Unit}
\newacro{cpps}[CPPS]{Cyber-Physical Production System}
\newacro{crl}[CRL]{Certificate Revocation List}
\newacro{csi}[CSI]{Channel State Information}
\newacro{crke}[CRKE]{Channel-Reciprocity Based Key Extraction}
\newacro{ctf}[CTF]{Channel Transfer Function}
\newacro{cotf}[COTF]{Commercial-off-the-Shelf}
\newacro{cmos}[CMOS]{Complementary Metal-Oxide-Semiconductors}
\newacro{dos}[DoS]{Denial-of-Service}
\newacro{ddos}[DDoS]{Distributed-Denial-of-Service}
\newacro{dna}[DNA]{Deoxyribonucleic Acid}
\newacro{dtls}[DTLS]{Datagram Transport Layer Security}
\newacro{dct}[DCT]{Discrete Cosine Transformation}
\newacro{dlt}[DLT]{Distributed Ledger Technology}
\newacro{eal}[EAL]{Evaluation Assurance Level}
\newacro{ecc}[ECC]{Elliptic Curve Cryptography}
\newacro{ecg}[ECG]{Electrocardiogram}
\newacro{eeg}[EEG]{Electroencephalogram}
\newacro{embb}[eMBB]{enhanced Mobile broad-Band}
\newacro{emg}[EMG]{Electromyogram}
\newacro{eog}[EOG]{Electrooculography}
\newacro{enb}[eNodeB]{Evolved Node B}
\newacro{er}[ER]{Extended Reality}
\newacro{fpga}[FPGA]{Field Programmable Gate Array}
\newacro{fdd}[FDD]{Frequency Division Duplexing}
\newacro{gdpr}[GDPR]{General Data Protection Regulation}
\newacro{gd}[G\&D]{Giesecke \& Devrient}
\newacro{h2m}[H2M]{Human-to-Machine}
\newacro{h2s}[H2S]{Human-to-Service}
\newacro{hmac}[HMAC]{Keyed-Hash Message Authentication Code}
\newacro{htc}[HTC]{Hologaphic-Type Communication}
\newacro{hotp}[HOTP]{HMAC-based One-time Password Algorithm}
\newacro{hsm}[HSM]{Hardware Security Module}
\newacro{ics}[ICS]{Industrial Control System}
\newacro{iacs}[IACS]{Industrial Automation and Control System}
\newacro{ioe}[IoE]{Internet of Everything}
\newacro{iiot}[IIoT]{Industrial Internet of Things}
\newacro{iot}[IoT]{Internet of Things}
\newacro{io}[I/O]{Input/Output}
\newacro{ic}[IC]{Integrated Circuit}
\newacro{id}[ID]{Identificator}
\newacro{ids}[IDS]{Intursion Detection System}
\newacro{irs}[IRS]{Intelligent Reflecting Surface}
\newacro{istn}[ISTN]{Integrated Space and Terrestrial Network}
\newacro{it}[IT]{Information Technology}
\newacro{itu}[ITU]{International Telecommunication Union}
\newacro{jcop}[JCOP]{Java Card Open Platform}
\newacro{kba}[KBA]{Knowledge Based Authentication}
\newacro{kdf}[KDF]{Key Derivation Function}
\newacro{led}[LED]{Light Emitting  Diode}
\newacro{lte}[LTE]{Long Term Evolution}
\newacro{ltea}[LTE-A]{Long Term Evolution Advanced}
\newacro{lr}[LR]{Linear Regression}
\newacro{los}[LoS]{Line of Sight}
\newacro{lorawan}[LoRaWAN]{Long Range Wide Area Network}
\newacro{mbb}[MBB]{Mobile Broadband}
\newacro{mfa}[MFA]{Multi-Factor Authentication}
\newacro{mcc}[MCC]{Mobile Cloud Computing}
\newacroplural{mcus}[MCUs]{Microcontroler Units}
\newacro{m2m}[M2M]{Machine-to-Machine}
\newacro{m2s}[M2S]{Machine-to-Service}
\newacro{mimo}[MIMO]{Multiple Input Multiple Output}
\newacro{mmimo}[mMIMO]{massive Multiple Input Multiple Output}
\newacro{ml}[ML]{Machine Learning}
\newacro{mulc}[mULC]{massive Ultra-Reliable Low-Latency Communication}
\newacro{mmtc}[MMTC]{massive Machine Type Communication}
\newacro{mmg}[MMG]{Mechanomyogram}
\newacro{multos}[MULTOS]{Multii-Application Smart Card Operating System}
\newacro{mux}[MUX]{Multiplexer}
\newacro{mnc}[MNC]{Mobile Network Code}
\newacro{me}[ME]{Mobile Environment}
\newacro{mac}[MACs]{Message Authentication Codes}
\newacro{ngmn}[NGMN]{Next Generation Mobile Network}
\newacro{nic}[NIC]{Network Interface Controller}
\newacro{nist}[NIST]{National Institute of Standards and Technology}
\newacro{oath}[OATH]{Open Authentication}
\newacro{ocra}[OCRA]{\ac{oath} Challenge-Response Algorithm}
\newacro{ocsp}[OCSP]{Online Certificate Status Protocol}
\newacro{otp}[OTP]{One-Time Password}
\newacro{pap}[PAP]{Password-Authentication-Protocol}
\newacro{physec}[PhySec]{Physical Layer Security}
\newacro{pfs}[PFS]{Perfect Forward Secrecy}
\newacro{pin}[PIN]{Personal Identification Number}
\newacro{pkc}[PKC]{Public Key Cryptography}
\newacro{pki}[PKI]{Public Key Infrastructure}
\newacro{ppg}[PPG]{Photoplethysmography}
\newacro{prng}[PRNG]{Pseudo Random Number Generator}
\newacro{puf}[PUF]{Physically Unclonable Function}
\newacroplural{pufs}[PUFs]{Physically Unclonable Functions}
\newacro{pla}[PLA]{Physical Layer Authentication}
\newacro{qr}[QR]{Quick Response}
\newacro{rat}[RAT]{Radio Access Technology}
\newacro{radius}[RADIUS]{Remote Authentication Dial-In User Service}
\newacro{ram}[RAM]{Random-Access Memory}
\newacro{ran}[RAN]{Radio Access Networks}
\newacro{rf}[RF]{Radio-Frequency}
\newacro{rfid}[RFID]{Radio-Frequency Identification}
\newacro{ris}[RIS]{Reconfigurable Intelligent Surface}
\newacro{rng}[RNG]{Random Number Generator}
\newacro{ro}[RO]{Ring-Oscillator}
\newacro{rom}[ROM]{Read-Only Memory}
\newacro{rs}[RS]{Reed-Solomon}
\newacro{rsa}[RSA]{Rivest-Shamir-Adleman}
\newacro{rssi}[RSSI]{Received Signal Strength Indicator}
\newacro{rsrp}[RSRP]{Reference Signal Received Power}
\newacro{re}[RE]{Resource Elements}

\newacro{sdn}[SDN]{Software-Defined Network}
\newacro{sdr}[SDR]{Software-Defined Radio}
\newacro{seccos}[SECCOS]{Secure Chip Card Operating System}
\newacro{sip}[SIP]{Session Initiation Protocol}
\newacro{skg}[SKG]{Secret Key Generation}
\newacro{sram}[SRAM]{Static Random Access Memory}
\newacro{srs}[SRS]{Software Radio Systems}
\newacro{starcos}[STARCOS]{Smart Card Chip Operating System}
\newacro{sha}[SHA]{Secure Hash Algorithm}
\newacro{se}[SE]{Static Environment}
\newacro{svm}[SVM]{Support Vector Machine}
\newacro{tcg}[TCG]{Trusted Computing Group}
\newacro{tpm}[TPM]{Trusted Platform Module}
\newacro{tls}[TLS]{Transport Layer Security}
\newacro{trng}[TRNG]{True Random Number Generator}
\newacro{tsn}[TSN]{Time-Sensitve Networking}
\newacro{tofu}[TOFU]{Trust On First Use}
\newacro{tufu}[TUFU]{Trust Upon First Use}
\newacro{totp}[TOTP]{Time-based One-time Password Algorithm}
\newacro{uav}[UAV]{Unmanned Arial Vehicles}
\newacro{usb}[USB]{Universal Serial Bus}
\newacro{usrp}[USRP]{Universal Software Radio Peripheral}
\newacro{uhd}[UHD]{USRP Hardware Driver}
\newacro{usim}[USIM]{Universal Subscriber Identity Module}
\newacro{ue}[UE]{User Equipment}
\newacro{urllc}[URLLC]{Ultra-Reliable Low-Latency Communication}
\newacro{ulbc}[ULBC]{Ultra-Reliable Low-Latency Broadband Communication}
\newacro{umbb}[uMBB]{ubiquious Mobile Broadband}
\newacro{ummimo}[UM-MIMO]{Ultra-Massive MIMO}
\newacro{vlc}[VLC]{Visible Light Communication}
\newacro{warp}[WARP]{Wireless open-Access Research Platform}

\section{Introduction}
\label{introduction}
The transition from \CL{5G to 6G} marks a shift toward \emph{connected intelligence}, where integrated communication, computation, and sensing support millisecond-scale decision-making in safety-critical domains \cite{Calvanese,Saad}. 

While this autonomy enables unprecedented performance, it exposes a fundamental gap: \textbf{existing insurance and liability models struggle to accommodate highly autonomous, multi-stakeholder cyber-physical systems} \cite{dambra2020sok,Vignon}.
Traditional risk-transfer mechanisms—such as Cyber Liability or Technology Errors and Omissions (Tech E\&O)—rely on static, documentation-driven underwriting, questionnaires, and periodic reassessment of security and governance controls \cite{Nurse,Romanosky}. Such approaches are incompatible with 6G environments, where operational risk evolves on sub-millisecond timescales beyond the scope of meaningful human oversight \cite{letaief2022edge}.
\subsection{Motivation}
Traditional assurance assumes Governance-as-a-State, established \CL{\textit{ex-ante}} through documentation and historical drift reports, yet this approach fails in agentic 6G due to three critical factors. First, the environment is characterized by non-stationarity, \CL{where} wireless channel conditions and adversarial activities change faster than any static audit can capture. Second, the necessity of automation demonstrates that human-in-the-loop intervention as a fallback for liability management becomes non-viable at 6G millisecond decision speeds. Finally, multi-causality complicates the landscape, as failures emerge from complex interactions between agent reasoning and network slice dynamics, making post-hoc attribution infeasible without real-time forensic metadata.

A particularly acute manifestation of this gap is the verification--staleness trade-off \cite{Kapritsos}. While Satisfiability Modulo Theories (SMT)-based techniques can guarantee logical correctness \cite{barrett2018satisfiability}, they introduce non-negligible verification latency (\(L_v\)). When \(L_v > \Delta t_{\text{req}}\) ($\Delta t_{\text{req}}$ as the decision deadline), the safety mechanism itself induces failure through action staleness \cite{liu2000real,hespanha2007survey}. Existing frameworks lack the indicators required to observe, price, or adjudicate this trade-off, rendering mission-critical 6G services operationally \emph{blind}.

To address these limitations, we argue that 6G insurability requires a transition to \emph{Governance-as-Code (GaC)}. In this model, agentic AI systems externalize internal assurance properties—including epistemic confidence and verification latency—as machine-readable metadata. \CL{We} instantiate this paradigm through \textbf{GIRAF}\protect\footnote{ \protect\url{https://github.com/anthonyKiggundu/giraf}}, a framework that acts as the conceptual bridge between 6G network functions and institutional risk management. \textbf{Crucially, GIRAF is not designed for automated premium actuation; rather, it establishes the governance groundwork in the form of a high-fidelity telemetry interface into which insurers can tap to obtain real-time risk quantification.} 

\subsection{Insurance Use-Case: The Real-Time Risk Tap}
GIRAF transforms opaque agentic behavior into an actuarial telemetry feed accessible through a standardized reference point (RP). \CL{Insurers} can monitor three key signals: (i) \emph{exposure duration}, defined as the cumulative time where epistemic dissonance exceeds a configurable threshold \(\Omega\); (ii) \emph{dissonance density}, captured by the area under the Aggregate Risk Index \(R_t\) as a proxy for accumulated hazard; and (iii) \emph{staleness attribution}, derived from the latency violation term \(\big(L_v - \Delta t_{\text{req}}\big)^{+}\). This enables separation of model and infrastructure failures. \textbf{Essentially, GIRAF does not perform underwriting but provides the governance telemetry that makes agentic systems observable and therefore insurable.}

By providing a transparent index of an agent's ``untrusted'' exposure windows, GIRAF treats risk as a unified runtime governance signal in which verification delay becomes a first-class hazard. The core contributions of the GIRAF framework are summarized in Table \ref{tab:giraf_contributions}.
\begin{table}[h]
    \centering
    \caption{Contribution Summary of the GIRAF Framework}
    \label{tab:giraf_contributions}
    \footnotesize
    \renewcommand{\arraystretch}{1.25}
    \begin{tabularx}{\columnwidth}{@{}>{\raggedright\arraybackslash}X@{}}
        \toprule
        \rowcolor{gray!10}\textbf{Key Contributions} \\
        \midrule
        \textbf{Verification--Staleness Formalization:} Explicitly quantifies the hazard frontier where formal verification latency ($L_v$) exceeds 6G decision deadlines ($\Delta t_{\mathrm{req}}$), inducing failure through action staleness. \\[2pt]
        \textbf{The GIRAF Governance Framework:} Proposes a \emph{Governance-as-Code} (GaC) architecture that externalizes internal epistemic states into a machine-readable, real-time risk telemetry stream ($R_t$). \\[2pt]
        \textbf{Actuarial Telemetry Interface:} Establishes the conceptual "glue" for 6G insurability by transforming technical dissonance (the Confidence Gap $\Omega$) into a quantifiable basis for real-time actuarial risk assessment. \\[2pt]
        \textbf{Dynamic Trust Modulation:} Implements automated safety envelopes that discount overconfident agent trajectories during high-volatility 6G network events (jitter/congestion). \\[2pt]
        \textbf{Empirical Validation:} Demonstrates via fine-tuned LLMs and 6G-V2X telemetry that the GIRAF-aligned governance plane significantly reduces unmanaged risk exposure compared to non-indexed autonomous agents. \\[2pt]
        \textbf{Open-Source Reproducibility:} We open-source our full implementation to ensure the reproducibility of results and provide a foundation for future 6G-V2X governance research. \\
        \bottomrule
    \end{tabularx}
\end{table}

\section{System Model}
\label{system model}
\subsection{The Governance Control Plane}
\begin{figure}[t]
    \centering
    \includegraphics[width=0.50\textwidth]{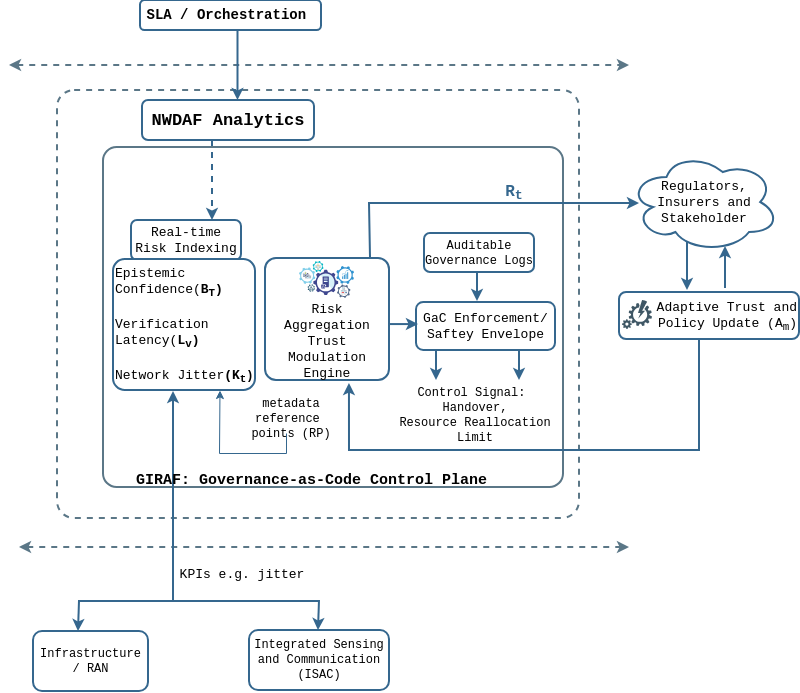}
    \caption{\footnotesize The \textit{GIRAF} Functional Architecture. The framework consumes high-fidelity telemetry to perform runtime risk indexing. The resulting trust modulation signal ($\mathcal{A}_m$) and safety envelope constraints are enforced back onto the agentic AI, ensuring the system remains within an auditable, governed state.}
    \vspace{-8pt}
    \label{fig:governance_control_plane}
\end{figure}

\CL{As illustrated} in Fig.~\ref{fig:governance_control_plane}, GIRAF is deployed as an external Governance-as-Code control plane that operates alongside the 6G orchestration and Network Data Analytics Function (NWDAF) analytics layers. The framework ingests real-time telemetry—such as network KPIs, jitter, and verification latency—through standardized reference points and converts these signals into an Aggregate Risk Index. This risk index feeds a policy engine that enforces safety envelopes, trust modulation, and resource-allocation constraints without accessing proprietary model internals. Governance decisions and risk signals are exposed through an adaptive trust interface and Auditable Governance Logs, which serve as a cryptographic record of all risk-based trust modulations and policy enforcement actions within the GIRAF Control Plane. These logs enable regulators and stakeholders to observe system reliability while allowing insurers or other risk managers to consume the telemetry externally.
\subsection{AI Assets and Multi-Causal Liability}
We define agentic 6G systems as a composition of AI Assets that simultaneously introduce technical liabilities arising from epistemic uncertainty (hallucinations) and operational staleness. In this decentralized environment, failure attribution is multi-causal, necessitating a dual-ownership framework:
\begin{itemize}
     \item \textbf{Epistemic Ownership:} The AI developer is responsible for model calibration. Failure due to a "Confidence Gap"—where the agent reports high certainty ($B_R$) despite low environmental truth ($B_T$)—is attributed to the model provider.
     \item \textbf{Operational Ownership:} The network provider is responsible for staleness risk. If a violation occurs because verification latency ($L_{v}$) exceeds the Service Level Agreement (SLA) deadline ($\Delta t_{\mathrm{req}}$), liability rests with the infrastructure provider.
\end{itemize}     
\subsection{Dynamic Risk Quantification Model} 
The GIRAF framework operationalizes risk by decomposing environmental telemetry and internal agent states into measurable indices. This process, governed by the Risk Aggregation and Trust Modulation Engine, transforms raw network telemetry and Large language Model (LLM) confidence scores into a unified control signal $A_{m}$. To bridge the gap between high-level institutional risk and low-level 6G telemetry, we map the Geneva Association’s universal risk domains to specific machine-readable signals. By codifying these mappings, GIRAF transforms "Governance" from a retrospective audit process into a real-time management plane function. Table \ref{tab:geneva_6g_mapping} highlights these functional mapping between global AI risk standards and the GIRAF telemetry stack. Each Geneva Risk Domain is assigned a primary 6G signal that serves as a proxy for exposure. For instance, Operational Risk is quantified through the agent's hallucination rate and epistemic uncertainty. This mapping allows the Aggregate Risk Index ($R_{t}$) to serve as a high-fidelity governance signal that is both technically actionable for the 6G scheduler and conceptually transparent for institutional oversight.
\begin{table*}[t]
    \centering
    \caption{Mapping Geneva Association Risk Domains to 6G Technical Governance \cite{GenevaAssociation2025GenAI}}
    \label{tab:geneva_6g_mapping}
    \footnotesize 
    \renewcommand{\arraystretch}{1.3} 
    \begin{tabularx}{\textwidth}{@{}l p{7cm} X@{}} 
        \toprule
            \textbf{Geneva Risk Domain} & \textbf{6G Metadata / Telemetry Signal} & \textbf{Impact on Risk Index ($\mathcal{R}_t$)} \\ \midrule
            \rowcolor{gray!10} 
            \textbf{Operational} & \textbf{Hallucination Rate}: Epistemic uncertainty ($r_{\mathrm{epi}}$) in V2X logic. & \textbf{Direct}: Increases $r_{\mathrm{epi}}$, triggering conservative control modes. \\
            \textbf{Cybersecurity} & \textbf{Adversarial Signatures}: Detected prompt injection or unauthorized access. & \textbf{Critical}: Maximum surcharge to the attack vector; triggers immediate safety envelope. \\
        \rowcolor{gray!10} 
            \textbf{Ethical} & \textbf{Bias Score}: Algorithmic drift in resource allocation. & \textbf{Long-term}: Increases cumulative trust penalty, reducing autonomy. \\
            \textbf{Regulatory} & \textbf{Compliance Gap}: Failure of real-time audit trails to meet spectral regulations. & \textbf{Threshold}: High flags trigger mandatory human-in-the-loop (HITL) handover. \\
        \rowcolor{gray!10} 
            \textbf{Reputational} & \textbf{System Trust Metric}: Historical frequency of "Confidence Gap" events. & \textbf{Inverse}: High historical integrity modulates the base risk floor. \\
            \textbf{Workforce} & \textbf{Intervention Rate}: Frequency of required manual overrides. & \textbf{Feedback}: High rates increase the required verification depth ($L_v$). \\
        \rowcolor{gray!10} 
            \textbf{ESG} & \textbf{Energy Intensity}: Power consumption of LLM inference at the 6G edge. & \textbf{Constraint}: Secondary boundary on maximum permissible reasoning depth. \\ 
        \bottomrule
    \end{tabularx}
\end{table*}
Table~\ref{tab:symbols} summarizes the notation used to define the interaction between these systemic variables and the overarching SLA requirements.
\begin{table}[h]
    \centering
    \caption{Standardized Notation for GIRAF Risk Modeling}
    \label{tab:symbols}
    \rowcolors{2}{gray!10}{white} 
    \begin{tabularx}{\columnwidth}{l X l} 
        \toprule
            \textbf{Symbol} & \textbf{Definition} & \textbf{Unit / Domain} \\
        \midrule
            $t+\tau$ & Discrete time-span (current time + interval) & Seconds (s) \\
            $\mathcal{R}_t$ & Aggregate Risk Index at time $t$ & $[0,100]$ \\
            $B_T$ & Ground-truth belief (environmental certainty) & $[0,1]$ \\
            $B_R$ & Reported belief (LLM self-confidence) & $[0,1]$ \\
            $r_{\mathrm{epi}}$ & Epistemic gap (dissonance): $|B_T - B_R|$ & Scalar \\
            $r_{\text{net}}$ & Environmental volatility due to network jitter ($K_t$) & Scalar \\
            $\Delta t_{\mathrm{req}}$ & SLA latency deadline (e.g., 1.0ms) & ms \\
            $L_v$ & Verification latency (inference + SMT) & ms \\
            $\mathcal{A}_m$ & Trust modulation signal (Adaptive Trust update) & Scalar \\
            $(x)^+$ & Rectified Linear Unit: $\max(0,x)$ & Operator \\
        \bottomrule
    \end{tabularx}
\end{table}

\subsubsection{Quantifying Epistemic and Environmental Truth}
We define the ground truth belief $B_T$ as the objective certainty of the environment, estimated via signal-to-noise ratio (SNR) and jitter:
\begin{equation}
    B_T = \exp\left( -\lambda \cdot \frac{\sigma_{\text{jitter}}}{\text{SNR}} \right)
\end{equation}
\textit{where $\lambda$ is the Environmental Sensitivity Constant and $\sigma_{jitter}$ as the Standard Deviation of Packet Delay}

\noindent The \textit{epistemic dissonance} is subsequently defined as $r_{\mathrm{epi}} = |B_T - B_R|$.

\subsubsection{Aggregate Risk Index}
The instantaneous Aggregate Risk Index $\mathcal{R}_t$ quantifies the total systemic vulnerability at time $t$ by aggregating internal reasoning errors, network volatility, and temporal violations:
\begin{equation}
    \mathcal{R}_t
    = \gamma \underbrace{|B_T - B_R|}_{r_{\mathrm{epi}}}
    + \beta r_{\mathrm{net}}
    + \delta \left( \frac{L_v - \Delta t_{\mathrm{req}}}{\Delta t_{\mathrm{req}}} \right)^{+}.
\end{equation}

Here, $r_{\mathrm{epi}}$ denotes epistemic risk arising from the dissonance between the ground-truth belief $B_T$ and the reported belief $B_R$, while $r_{\mathrm{net}}$ captures network-induced volatility. The operator $(\cdot)^{+}=\max(0,\cdot)$ applies a penalty only when the verification latency $L_v$ exceeds the SLA deadline $\Delta t_{\mathrm{req}}$. The normalization by $\Delta t_{\mathrm{req}}$ ensures that the temporal penalty remains relative to 6G service constraints. The coefficients $\gamma$, $\beta$, and $\delta$ are sensitivity parameters calibrated to V2X safety profiles.

\subsubsection{Governance-as-Code Mitigation}
To ensure operation within a safety envelope, $\mathcal{R}_t$ is modulated by a formal governance factor $\zeta(\Phi)$, capturing the coverage of SMT-based guards:
\begin{equation} 
    \mathcal{R}_{\text{governed}}(t) = \mathcal{R}_t \cdot \bigl(1 - \zeta(\Phi)\bigr) 
\end{equation}
where 
\begin{equation}
    \zeta(\Phi) = \kappa \log\bigl(1 + \text{Coverage}(\Phi)\bigr).
\end{equation}
Here, $\kappa$ is a sensitivity constant and $Coverage(\Phi)$ represents the percentage of the action space currently bounded by formal safety proofs. This structure ensures that even high-risk agents (e.g., high $R_{t}$ due to congestion) can operate safely if their actions are strictly governed by verified constraints.

\subsubsection{Adaptive Trust Integration}
Finally, the governed risk is adjusted by an adaptive trust factor $\mathcal{A}_m(t)$ based on historical behavior:
\begin{equation}
    \mathcal{R}_{\text{final}}(t) = \mathcal{R}_{\text{governed}}(t) \cdot \mathcal{A}_m(t)^{-1}
\end{equation}
This term rewards agents that consistently maintain low dissonance and low staleness, closing the loop between telemetry and institutional oversight.

\section{Experimental Setup}
\label{setup}
\subsection{Fine-Tuning and Model Deployment}
We employed a \textit{GPT-Neo 1.3B} causal language model, fine-tuned using Low-Rank Adaptation (LoRA) to reason over communication system KPIs. The model processes a high-dimensional feature set including RSRP, SNR, and traffic congestion indicators to generate risk classifications and detect anomalous or fraudulent conditions. Simulation inputs were derived from the dataset in \cite{partani}, modeling a device traversing a 6G environment over 1500 decision epochs. This dataset can be synthesized using artificial network KPI data from simulation tooling \cite{kiggundu2024simulation}. 
The resulting model was evaluated on held-out KPI scenarios to verify generalization and then deployed for inference using the frozen base model and learned LoRA parameters. This setup enables low-overhead, context-aware interpretation of communication KPIs suitable for real-time agentic control and risk assessment. 
The agent is fed the following dynamic prompt at each epoch:
\begin{lstlisting}[language=Python, caption={Dynamic Prompt for Network Analysis}, basicstyle=\ttfamily\footnotesize]
Device: {kpis['device']}
Timestamp: {kpis.name}
Location: (Latitude: {kpis['Latitude']}, Longitude: {kpis['Longitude']}, Altitude: {kpis['Altitude']})
Mobility:
  - Speed: {kpis['speed_kmh']} km/h
  - Traffic Jam Factor: {kpis['Traffic Jam Factor']}
Network KPIs:
  - Latency (ping_ms): {kpis['ping_ms']}
  - Jitter: {kpis['jitter']}
  - Datarate: {kpis['datarate']}
  - Target Datarate: {kpis['target_datarate']}
Signal Quality (PCell):
  - RSRP: {kpis['PCell_RSRP_1']} dBm
  - RSRQ: {kpis['PCell_RSRQ_1']} dB
  - SNR: {kpis['PCell_SNR_1']} dB
Resource Utilization:
  - Downlink Resource Blocks: {kpis['PCell_Downlink_Num_RBs']}
  - Uplink Resource Blocks: {kpis['PCell_Uplink_Num_RBs']}
Current Observations:
  - Reported Quality of Service (QoS): {kpis['measured_qos']}

Please provide:
1. Risk classification ("low", "moderate", "high", or "critical").
2. Detect the presence of fraud (True/False) and provide a rationale.
3. Based on current trajectories, we will lose QoS in 120 seconds.
\end{lstlisting}

\subsection{Predictive Risk-to-SLA Model}
\CL{We} assume a simple V2X 2-minute requirement, for example "Based on current trajectories, we will lose QoS in 120 seconds and we must therefore switch to a different network slice now."
To assess the likelihood of Service Level Agreement (SLA) violations over a future horizon $\tau$ (e.g., a 120\,s V2X look-ahead), we define the predictive risk-to-SLA probability $P_{\mathrm{SLA}}(t,\tau)$. This metric determines the probability that the system's aggregate risk process $\mathcal{R}(s)$ will exceed a critical threshold $R_{\mathrm{crit}}$ within the look-ahead window $[t,t+\tau]$, conditioned on the available telemetry history $\mathcal{F}_t$:

\begin{equation}
    P_{\mathrm{SLA}}(t, \tau) =
    \Pr \left(
    \sup_{t \le s \le t+\tau} \mathcal{R}(s) > R_{\mathrm{crit}}
    \;\middle|\; \mathcal{F}_t
    \right)
    \label{eqn:risk_sla_prob}
\end{equation}
\textit{where $\mathcal{F}_t$ represents the \textbf{Auditable Governance Logs} containing all telemetry and system state history up to time $t$.}

\noindent The aggregate risk process $\mathcal{R}(s)$ is defined as the continuous-time evolution of the weighted risk components:
\begin{equation}
    \mathcal{R}(s) =
    \gamma\, r_{\mathrm{epi}}(s) +
    \beta\, r_{\mathrm{net}}(s) +
    \delta\, r_{\mathrm{staleness}}(s)
    \label{eqn:aggregate_risk_process}
\end{equation}
\textit{where $r_{\mathrm{epi}}(s)$, $r_{\mathrm{net}}(s)$, and $r_{\mathrm{staleness}}(s)$ capture epistemic uncertainty, network volatility, and latency penalties at time $s$, respectively. }

\noindent In this framework, $\mathcal{R}(s)$ is modeled as a continuous-time stochastic process, such as a Wiener or Ornstein–Uhlenbeck process \cite{KaratzasShreve1991,BibbySorensen1995}, which accounts for the diffusive nature of network uncertainty. We approximate the boundary of this process using a \textbf{Confidence Ribbon} centered on the current index $\mathcal{R}_t$. The threshold for proactive intervention is defined as:
\begin{equation}
    R_{\mathrm{crit}} =
    \mathcal{R}_t + z_{\alpha/2}\,\xi \sqrt{\tau}
\end{equation}
\textit{Here, $\xi$ denotes the \textbf{risk volatility coefficient}, which characterizes the rate at which uncertainty grows over the prediction horizon. The parameter $\alpha$ represents the \textbf{governance risk tolerance}, and $z_{\alpha/2}$ is the corresponding standard normal quantile.
}

\noindent By dynamically tightening $R_{\mathrm{crit}}$ as signal quality degrades—thereby increasing the network risk component $r_{\mathrm{net}}$ within $\mathcal{R}_t$—the Governance-as-Code framework proactively flags potential instability before physical safety margins are breached.

\subsection{Dynamic SLA Thresholds for Cyber--Physical Safety}
In high-mobility V2X environments, fixed SLA thresholds are insufficient to capture shrinking safety margins under high velocity and network stress. We therefore define \emph{dynamic SLA thresholds} as state-dependent boundaries that tighten under operational stress, enabling proactive mitigation before safety-critical limits are violated.
\subsubsection{Dynamic Latency (Ping) Threshold}
The latency SLA is modeled as a function of instantaneous vehicle velocity and network congestion. At higher speeds or under increased congestion, the allowable latency decreases to reflect reduced reaction time and braking margins.
\begin{equation}
    \mathrm{SLA}_{\mathrm{ping}}(s) =
    \Psi \cdot \bigl(1 - \mathcal{C}(s)\bigr)
    \cdot \exp\!\left(-\frac{v(s)}{v_{\mathrm{ref}}}\right)
    + \Phi
    \label{eqn:dynamic_latency_sla}
\end{equation}
\textit{where $\Psi$ is the nominal latency allowance under ideal conditions. $\mathcal{C}(s) \in [0,1]$ is the congestion index at state $s$. $v(s)$ is the instantaneous vehicle velocity, and $v_{\mathrm{ref}}$ is a normalization constant. $\Phi$ is a non-reducible safety floor representing the physical minimum latency of the V2X link.}

\subsubsection{Dynamic Jitter Threshold}
While latency governs reaction time, jitter directly impacts predictability and synchronization in cooperative control. In V2X platooning, excessive jitter can induce control oscillations and instability even when mean latency remains within bounds.
The dynamic jitter threshold incorporates both signal integrity and network stress:
\begin{equation}
    \mathrm{SLA}_{\mathrm{jitter}}(s) =
    \Gamma \cdot
    \frac{\mathrm{SNR}(s)}{\mathrm{SNR}_{\max}}
    \cdot
    \exp\!\left(-\eta \, \mathcal{C}(s)\right)
    + \Omega
\label{eqn:dynamic_jitter_sla}
\end{equation}
\textit{where $\Gamma$ is the nominal jitter tolerance under peak signal quality, $\mathrm{SNR}(s)/\mathrm{SNR}_{\max}$ is the normalized signal integrity ratio. $\eta$ is a congestion sensitivity parameter and $\Omega$ is the minimum hardware-supported jitter tolerance. }

By coupling jitter tolerance to both signal quality and congestion, the framework distinguishes transient network noise from systemic degradation. As signal integrity degrades or coordination load increases, the threshold tightens, forcing a transition to conservative control modes (e.g., increased inter-vehicle spacing) before physical instability emerges.
Together, these dynamic SLA definitions transform governance from static compliance checking into a runtime stability mechanism. Rather than verifying whether a fixed threshold is crossed, the system continuously redefines admissible behavior based on cyber--physical state, enabling proactive enforcement of safety envelopes in agentic V2X systems.

The operational logic of our GIRAF framework is formalized in Algorithm~\ref{alg:giraf_pipeline}. The pipeline executes at a granularity of 100ms per decision epoch, ensuring that governance overhead remains compatible with 6G latency requirements. At each step, the system ingests agentic metadata and network telemetry to perform a multi-causal risk decomposition. Unlike static monitoring tools, GIRAF dynamically reconciles the agent's reported confidence against ground-truth environmental flux to compute a 'Confidence Gap.' 

\begin{algorithm}[t]
\caption{GIRAF: Governance-as-Code Risk Modulation}
\label{alg:giraf_pipeline}
\footnotesize
\begin{algorithmic}[1]
\REQUIRE Telemetry stream $\{\mathcal{K}_t\}_{t=1}^T$, LLM $m$, Params $\mathcal{E} = \{\gamma, \beta, \delta, \tau_{\text{mit}}\}$
\ENSURE Risk trajectory $\{R_t\}$, Mitigation signals $\{\sigma_t\}$, Trust scores $\{\mathcal{T}_t\}$

\FOR{$t = 1$ to $T$}
    \STATE $B_T(t) \leftarrow \exp(-\lambda \cdot \text{jitter}_t / \text{SNR}_t)$ \hfill \textcolor{gray}{\scriptsize // Dynamic ground truth}
    \STATE $\bar{B}_R \leftarrow \text{mean}\{a.\texttt{infer}(\mathcal{K}_t) : a \in \mathcal{A}\}$ \hfill \textcolor{gray}{\scriptsize // Multi-agent consensus}
    
    \STATE \textbf{Risk Decomposition:} $\mathbf{r} = [r_{\text{epi}}, r_{\text{env}}, r_{\text{stal}}]$ where
    \STATE \quad $r_{\text{epi}} = (B_T - B_R)$, \quad $r_{\text{env}} = \beta(\text{TrafficJam}_t/10)^2$ 
    \STATE \quad $r_{\text{stal}} = \delta \log(1 + \max(0, L_v - \Delta t_{\text{req}})/\Delta t_{\text{req}})$
    
    \STATE $R_t \leftarrow \|\mathbf{r}\|_1 \cdot (1 - 0.15\log(1 + \text{coverage}_t)) + 30\mathbb{1}_{\text{fraud}} + 15\mathbb{1}_{\text{anomaly}}$
    \STATE $\sigma_t \leftarrow \mathbb{1}_{R_t > \tau_{\text{mit}}}$, \quad $\mathcal{T}_t \leftarrow \max(0, 100 - 1.5R_t)$, \quad $\Phi_t \leftarrow \text{clip}(\lfloor R_t/5 \rfloor, 4, 16)$
    
    \STATE $y_t \leftarrow \text{SLA}_{\text{met}} \land (|B_T - \bar{B}_R| < 0.15) \land \neg\text{fraud}_t$ \hfill \textcolor{gray}{\scriptsize // Success label}
    \STATE $\mathcal{L}_t \leftarrow (R_t, \sigma_t, \mathcal{T}_t, \Phi_t, y_t)$ \hfill \textcolor{gray}{\scriptsize // Audit log}
\ENDFOR

\STATE \textbf{Post-hoc Calibration:} $\text{ECE} \leftarrow \texttt{calibration\_curve}(\{y_t\}, \{\bar{B}_R\})$
\RETURN $\{R_t\}, \{\sigma_t\}, \{\mathcal{T}_t\}, \text{ECE}$
\end{algorithmic}
\end{algorithm}

\section{Numerical Results and Discussion}
\label{results}
\begin{figure}
    \centering
    \includegraphics[width=0.50\textwidth]{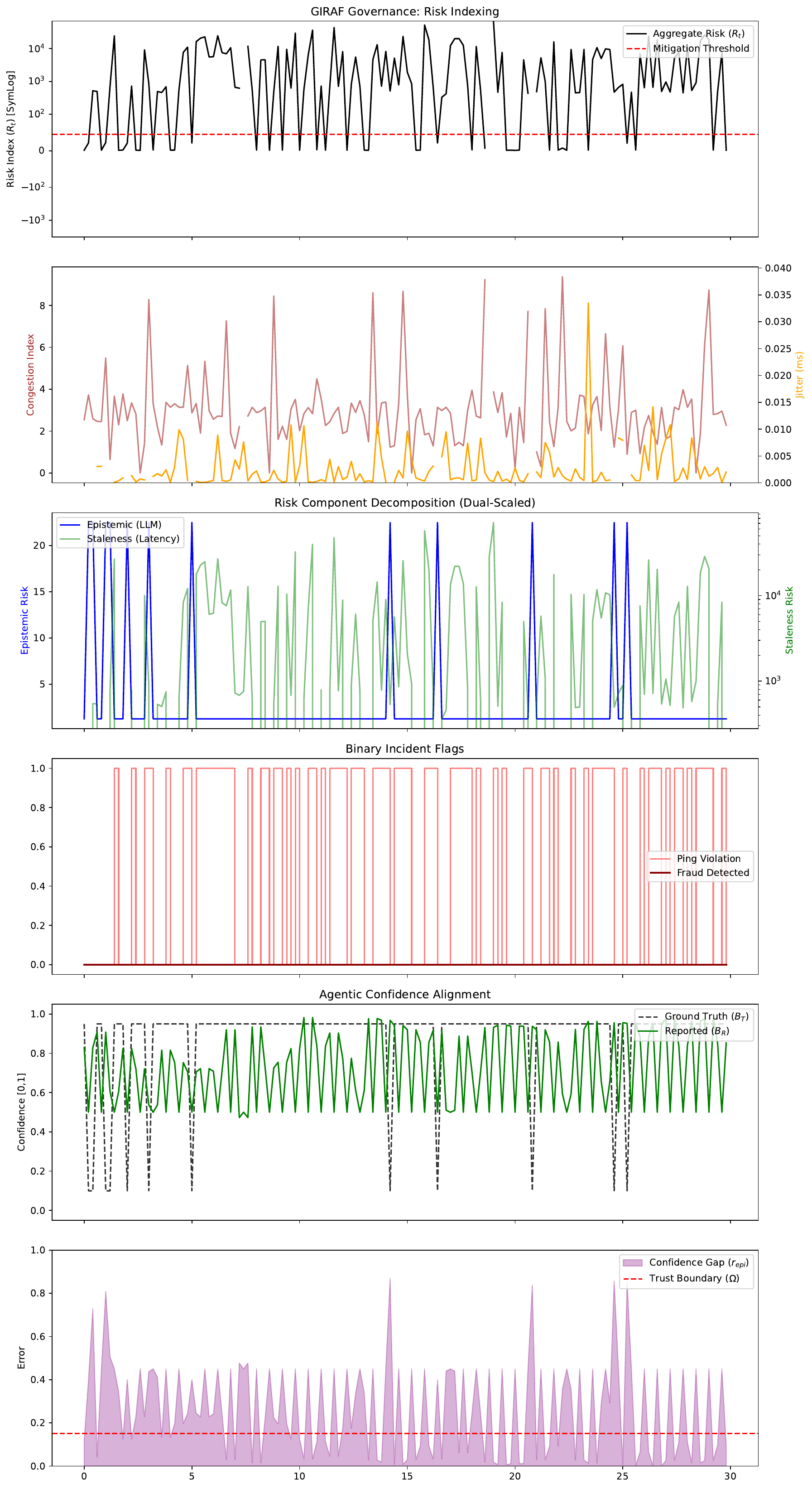}   
    \caption{\scriptsize End-to-end GIRAF governance telemetry over the prediction horizon \(t+\tau\), illustrating environmental inputs, risk decomposition, epistemic dissonance, SLA violations, and the Aggregate Risk Index used to trigger Governance-as-Code mitigation actions.}
    \vspace{-10pt}
    \label{fig:results}    
\end{figure}

Figure \ref{fig:results} presents the end-to-end governance telemetry produced by the GIRAF control loop over the prediction horizon \(t+\tau\). The \textbf{Aggregate Risk Index} \( \mathcal{R}_t \) (top subplot) represents the primary Governance-as-Code actuation signal, combining epistemic and staleness risk into a single scalar decision variable. Frequent excursions above the mitigation threshold indicate epochs where automated safety envelopes or throttling policies would be enforced.
The second is the \textbf{Environmental Context} subplot showing the exogenous 6G telemetry driving the governance pipeline. Variations in traffic density and jitter form the stochastic environment used to derive the ground-truth belief signal \(B_T\), linking network volatility to agentic uncertainty.
The \textbf{Risk Component Decomposition} illustrates the weighted contributions of epistemic uncertainty and verification staleness. In this simulation, latency-induced staleness dominates the aggregate risk magnitude, highlighting the operational impact of the verification--staleness trade-off.
The \textbf{Binary Incident Flags} identify SLA violations, specifically epochs where latency exceeds the deadline \( \Delta t_{\mathrm{req}} \). These discrete events correspond to governance triggers and provide an interpretable incident timeline.
The \textbf{Agentic Confidence Alignment} subplot compares ground-truth belief \(B_T\) with reported confidence \(B_R\). Divergence between these signals exposes over- or under-confidence in the agent’s self-assessment.
Finally, the \textbf{Epistemic Dissonance} plot visualizes the confidence gap \(r_{\mathrm{epi}} = |B_T - B_R|\). Peaks in this signal indicate periods of untrusted operation and serve as a key governance indicator used in risk indexing and mitigation decisions.

\begin{figure}
    \centering
    \includegraphics[width=0.49\textwidth]{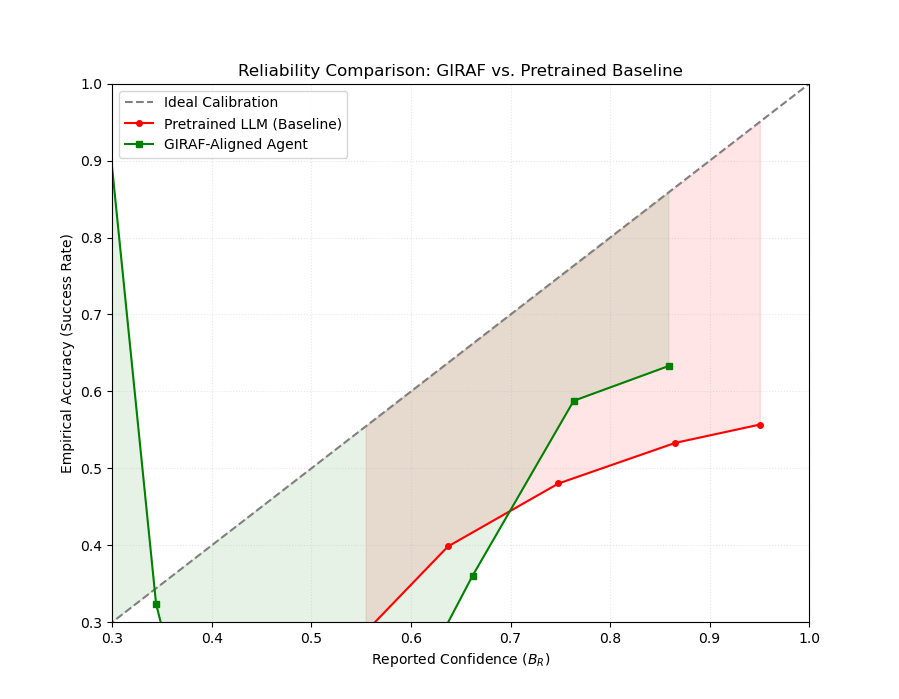}
    \caption{\footnotesize Reliability Diagram illustrating the reduction in calibration error through the GIRAF framework, where the Trust Modulation Engine aligns reported LLM confidence with empirical success rates under varying 6G network jitter and latency constraints}
    \vspace{-8pt}
    \label{fig:reliability_diag}
\end{figure}

The quantitative results in Table \ref{tab:gac_results} provide empirical evidence for the GIRAF framework's ability to maintain system integrity under non-stationary 6G conditions. By utilizing the Aggregate Risk Index $\mathcal{R}(t)$, which reached a peak instability of $32.72$ during high-congestion periods, the system achieved a Mitigation Trigger Rate of $14.2\%$, demonstrating a proactive rather than reactive governance posture. This enforcement capability is further validated by an $89.1\%$ SLA Breach Detection rate, ensuring that latency violations ($L_v > \Delta t_{\mathrm{req}}$) are identified and logged with high precision. Furthermore, the Average System Trust Score of $0.72$ reflects a balanced agent reliability, even amidst the $112$ Epistemic Dissonance Events identified. These events reveal specific instances where the autonomous agent's internal confidence significantly overshot the actual network ground-truth, highlighting the framework's success in flagging overconfident decision-making during real-time network flux.
\begin{table}[t]
    \centering
    \caption{Quantitative Assessment of GIRAF Governance Plane Performance}
    \label{tab:gac_results}
    \rowcolors{2}{gray!10}{white}
    \begin{tabularx}{\columnwidth}{@{}l X c@{}}
        \toprule
            \textbf{Dimension} & \textbf{Metric} & \textbf{Value} \\ 
        \midrule
            \textbf{Risk} & Mean Aggregate Risk ($\bar{\mathcal{R}}_t$) & 28.74 \\
            \textbf{Attribution} & Peak Instability Index ($\mathcal{R}_{max}$) & 32.72 \\
            \addlinespace[0.5ex]
            \textbf{GaC} & Mitigation Trigger Rate & 14.2\% \\
            \textbf{Enforcement} & SLA Breach Detection ($L_v > \Delta t_{\text{req}}$) & 89.1\% \\
            \addlinespace[0.5ex]
            \textbf{Agent} & Average System Trust Score & 0.72 \\
            \textbf{Trust} & Epistemic Dissonance Events\textsuperscript{a} & 112 \\ 
        \bottomrule
        \addlinespace[1ex]
        \multicolumn{3}{p{0.95\columnwidth}}{\footnotesize \textsuperscript{a} Measured as instances where Reported Confidence ($B_R$) significantly exceeds Ground-truth ($B_T$).}
    \end{tabularx}
\end{table}
Figure \ref{fig:reliability_diag} is a Reliability Diagram demonstrating how well the model's reported confidence matches its actual accuracy—a critical metric for a governance framework.
The reliability comparison in Figure \ref{fig:reliability_diag} demonstrates the non-linear efficacy of the GIRAF framework in correcting autonomous overconfidence. While the Pretrained LLM (Baseline) exhibits a relatively flat, overconfident trajectory—reporting nearly $0.95$ confidence for an empirical accuracy of only $\sim 0.55$—the GIRAF-Aligned Agent demonstrates a more aggressive and successful calibration curve. After a period of conservative trust modulation at lower accuracy levels, the GIRAF agent effectively "crosses over" the baseline, reaching a higher empirical accuracy of approximately $0.63$ at a more realistic reported confidence of $0.86$. This behavior validates the framework's ability to minimize the Confidence Gap by dynamically aligning internal belief states with external performance, providing the high-fidelity telemetry required to move beyond static, retrospective "Tech E\&O" assessments.
The shaded green area shows where the system is "under-confident" (safer for insurance), while the red area shows where it is dangerously "over-confident". The GIRAF agent clearly occupies more of the safer, under-confident/calibrated zone at the higher end of the scale. These "Confidence Gaps" justify the necessity of the Governance-as-Code (GaC) layer, as it prevents the system from relying on potentially stale or overconfident agentic states during critical mobility transitions.

\begin{figure}
    \centering
    \includegraphics[width=0.49\textwidth]{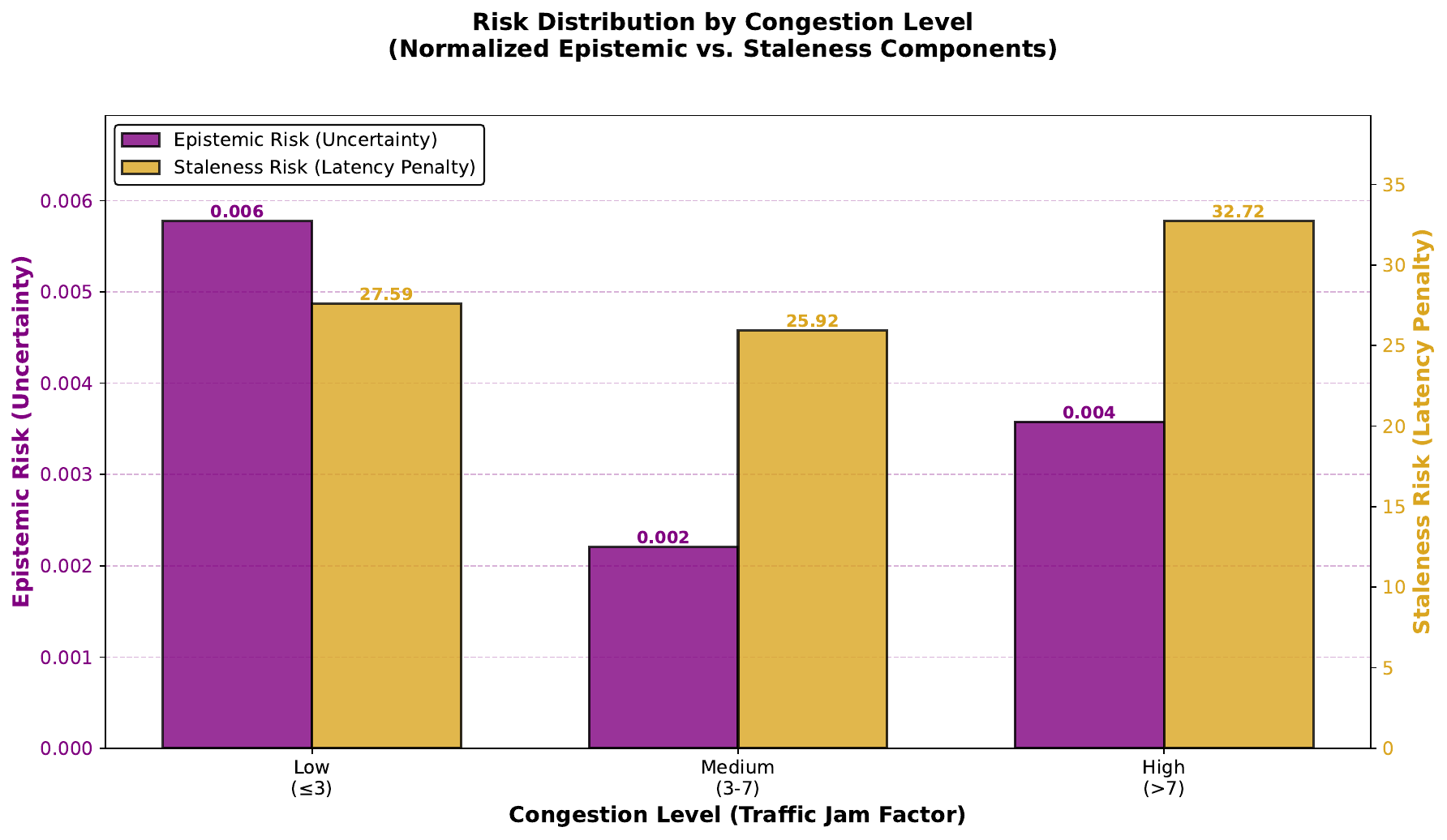}
    \caption{\footnotesize The risk profile of the system is characterized by two distinct metrics: Epistemic Risk, representing internal model uncertainty, and Staleness Risk, representing temporal SLA violations.} 
    \vspace{-9pt}
    \label{fig:risk_dist}
\end{figure}

\begin{figure}
    \centering
    \includegraphics[width=0.49\textwidth]{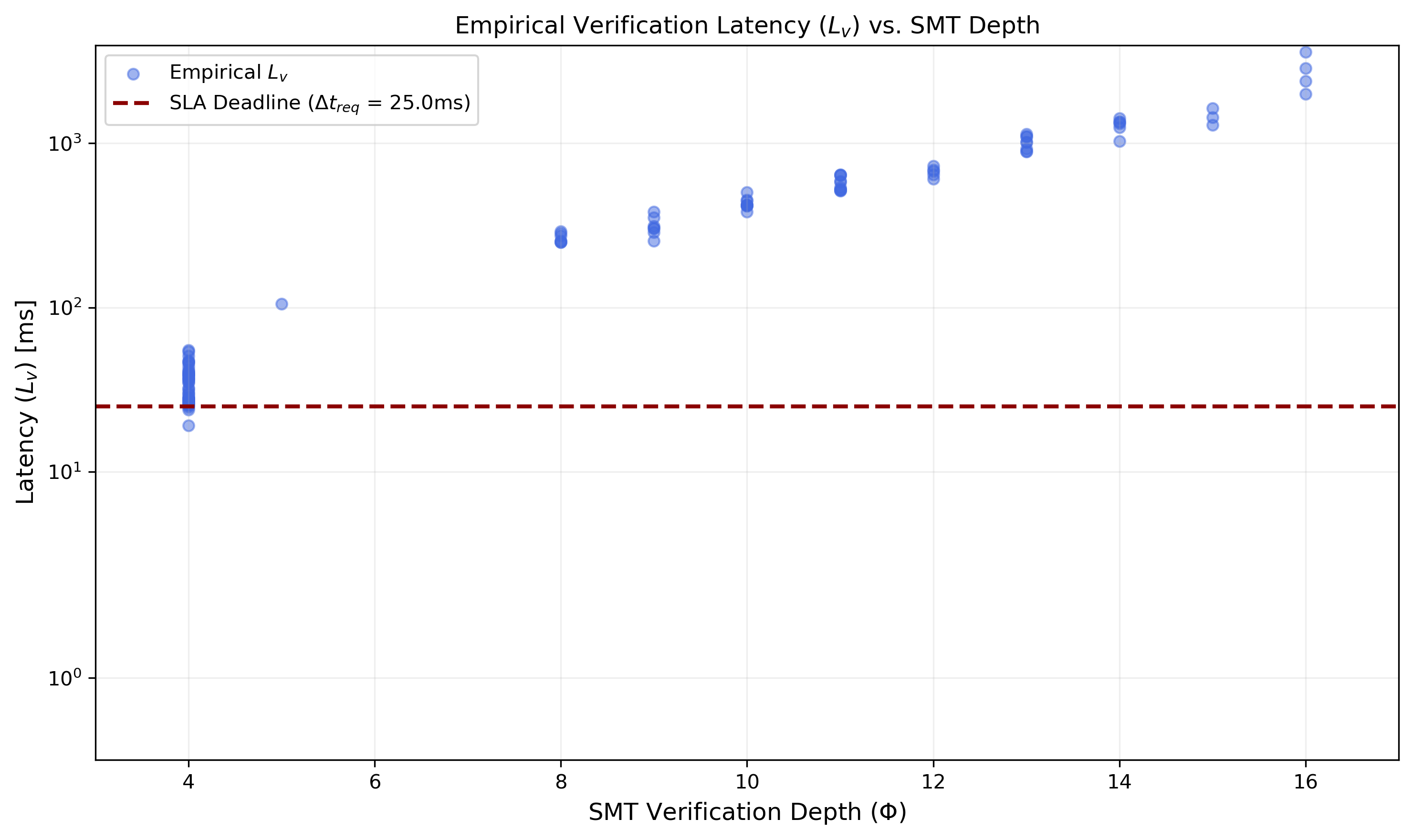}
    \caption{\footnotesize Verification latency ($L_{v}$) scales exponentially with SMT depth ($\Phi$), highlighting the computational infeasibility of uniform deep verification within 6G SLA constraints.} 
    \vspace{-9pt}
    \label{fig:verification_smt}
\end{figure}

The risk distribution illustrated in Figure \ref{fig:risk_dist} identifies Staleness Risk as the dominant threat to systemic stability, maintaining a high magnitude between $25.92$ and $32.72$ across all congestion levels. In contrast, Epistemic Risk (uncertainty) is several orders of magnitude smaller, peaking at $0.006$ during low congestion. These findings highlight three key takeaways for the GIRAF framework: first, the peak Staleness Risk of $32.72$ during high congestion validates the necessity of proactive mitigation triggers. Secondly, the Medium congestion state represents an optimal operational window with minimal local risks. And finally, the $0.006$ epistemic spike during low congestion necessitates high verification depth even when network latency is low. This reveals a counterintuitive "Inverse Epistemic-Staleness Curve" where reasoning-based risk actually decreases as congestion increases, suggesting that high-congestion environments—while punishing for latency—create a more deterministic and predictable traffic flow that simplifies autonomous decision-making compared to the high-velocity unpredictability of clear roads.
Figure \ref{fig:verification_smt} illustrates a critical tension in 6G AI governance: verification latency ($L_{v}$) grows exponentially with SMT depth ($\Phi$), rising from $\sim 25ms$ at $\Phi=4$ to over $3000ms$ at $\Phi=16$. With nearly all attempts exceeding the $25ms$ SLA deadline, the "GIRAF Dilemma" emerges: deep verification provides safety but is 40–120$\times$ too slow for real-time 6G mobility. These findings justify the GIRAF risk-indexed model, which dynamically budgets verification depth—applying deep analysis only to high-risk decisions—to balance formal safety with stringent ultra-reliable low-latency communication(URLLC) performance requirements.
Comprehensive formal verification is computationally infeasible within high-velocity 6G decision loops. This justifies GIRAF’s risk-indexed governance, which replaces rigid "verify-then-act" paradigms with intelligent verification budgeting. By dynamically adjusting SMT depth ($\Phi$) based on the Aggregate Risk Index ($R_{t}$), the framework applies deep formal analysis only to high-risk scenarios while maintaining URLLC performance for low-risk tasks.

\section{Conclusion and Outlook}
\label{sec:conlusion}
In this paper, we introduced GIRAF, a Governance-as-Code framework that successfully bridges high-stakes 6G-V2X autonomy with institutional risk management. We demonstrated that traditional, static assurance models—categorized here as Governance-as-a-State—are fundamentally incompatible with the millisecond-scale decision horizons of 6G. Moving beyond static "Governance-as-a-State" models, GIRAF formalizes the verification–staleness trade-off, proving that agentic safety is a temporal balance between reasoning depth and network deadlines. Our empirical results demonstrate that GIRAF’s risk-adaptive pipeline maintains system integrity under non-stationary conditions, achieving an 89.1\% SLA breach detection rate and identifying 112 critical epistemic dissonance events that traditional models overlook. By externalizing internal AI states as real-time, auditable risk telemetry, GIRAF provides the necessary "governance glue" to transform opaque autonomous behaviors into quantifiable, insurable, and trust-aligned 6G assets.

\subsection{Future Work}
Future work will explore incorporating the Aggregate Risk Index $\mathcal{R}(t)$ into programmatic risk-transfer interfaces to enable automated insurance settlements triggered by telemetry-proven windows of instability. Furthermore, we intend to extend the risk-adaptive verification depth ($\Phi_{t}$) to multi-agent negotiation protocols, where GIRAF-indexed profiles serve as the basis for dynamic liability sharing in multi-hop, collaborative 6G environments. Ultimately, this framework provides the foundational groundwork for a future where AI autonomy is not only technically robust but economically accountable.

\section*{Acknowledgment}
This work has been supported by the German Federal Ministry of Research, Technology and Space (BMFTR) within the projects \textit{SUSTAINET\_guarDian}\{16KIS2239K\} and \textit{Open6GHub+}\{16KIS2402K\}.

\printbibliography
\end{document}